\documentclass[conference]{IEEEtran}
\IEEEoverridecommandlockouts
\usepackage{cite}
\usepackage{amsmath,amssymb,amsfonts}
\usepackage{algorithm}
\usepackage{algorithmic}
\usepackage{graphicx}
\usepackage{textcomp}
\usepackage{xcolor}
\usepackage{stfloats}
\def\BibTeX{{\rm B\kern-.05em{\sc i\kern-.025em b}\kern-.08em
    T\kern-.1667em\lower.7ex\hbox{E}\kern-.125emX}}
\begin{document}

\title{Random Pilot and Data Access for Massive MIMO Spatially Correlated Rayleigh Fading Channels\\}
\author{\IEEEauthorblockN{Junyuan Gao, Yongpeng Wu, and Fan Wei}
\thanks{J. Gao, Y. Wu, and F. Wei are with the Department of Electronic Engineering, Shanghai Jiao Tong University, Minhang 200240, China (e-mail: sunflower0515@sjtu.edu.cn; yongpeng.wu@sjtu.edu.cn; weifan89@sjtu.edu.cn) (Corresponding author: Yongpeng Wu).}
}
\maketitle

\begin{abstract}
  Random access is necessary in crowded scenarios due to the limitation of pilot sequences and the intermittent pattern of device activity. Nowadays, most of the related works are based on independent and identically distributed (i.i.d.) channels. However, massive multiple-input multiple-output (MIMO) channels are not always i.i.d. in realistic outdoor wireless propagation environments.
  In this paper, a device grouping and pilot set allocation algorithm is proposed for the uplink massive MIMO systems over spatially correlated Rayleigh fading channels. Firstly, devices are divided into multiple groups, and the channel covariance matrixes of devices within the same group are approximately orthogonal. In each group, a dedicated pilot set is assigned. Then active devices perform random pilot and data access process. The mean square error of channel estimation (MSE-CE) and the spectral efficiency of this scheme are derived, and the MSE-CE can be minimized when collision devices have non-overlapping angle of arrival (AoA) intervals.
  Simulation results indicate that the MSE-CE and spectral efficiency of this protocol are improved compared with the traditional scheme. The MSE-CE of the proposed scheme is close to the theoretical lower bound over a wide signal-to-noise ratio (SNR) region  especially for long pilot sequence. Furthermore, the MSE-CE performance gains are significant in high SNR and strongly correlated scenarios.
\end{abstract}


\section{Introduction}
As the society is becoming fully networked, the number of wireless devices and the amount of data traffic are growing rapidly, which calls for the development of the fifth generation (5G) wireless communication \cite{b1}. It is known that 5G wireless networks will support three generic services including Enhanced Mobile BroadBand (eMBB), Massive Machine Type Communication (mMTC), and Ultra-Reliable Low Latency Communications (URLLC) \cite{b1}. Among them, mMTC aims to achieve the communications between large number of low-cost and sporadically active devices with low-data rate \cite{b2}. It has been a necessary service driven by many newly emerging use cases, such as Internet of Things (IoT) and machine-to-machine communications. Hence, the reliable support for massive connectivity of devices has been an important issue.

Channel state information (CSI) plays an important role in coherent communication. In time-division duplex (TDD) massive multiple-input multiple-output (MIMO) transmission, uplink (UL) CSI at the base station (BS) can be estimated through orthogonal pilots, and downlink channel estimates can be obtained utilizing channel reciprocity \cite{b3}. In massive access scenarios, the channel estimation is challenging because of two reasons \cite{b4}. First, since devices are generally low-cost, the duration of pilot sequences is limited by the uplink power budget. Once devices are mobile, it is also limited by the channel coherence time \cite{b5}. Hence, the number of devices is much larger than that of available orthogonal pilots, and it becomes impossible for devices to have dedicated pilots. Second, each device sends data to the BS in an intermittent pattern. Therefore, it is not necessary to allocate dedicated pilots to all the devices within the network \cite{b4}. These are the key motivations for the study of random access.

Nowadays, many works focus on random uplink access in massive MIMO \cite{b4,b5,b6,b7}.
For example, in \cite{b4}, each device is assigned with a unique non-orthogonal pilot hopping pattern. According to these patterns, active devices select pilots in training phases within multiple transmission slots and data codewords are transmitted afterwards. Hence, devices can be identified and their codewords can be merged.
In \cite{b6}, coded access and successive interference cancellation is used to realize random uplink access.
The strongest-user collision resolution decision criterion was proposed in \cite{b7}. Each active device randomly selects a pilot from the pilot set, but only the device with the strongest path-gain can access the network successfully.
However, in these literatures, only independent and identically distributed (i.i.d.) channels are considered. In realistic outdoor wireless propagation environments, the BS is located at an elevated position and the scattering around the BS is limited. Hence, most of the channel power lies in a finite number of spatial directions \cite{b8}.

In this paper, we consider random pilot and data access in massive MIMO systems with spatially correlated Rayleigh fading channels. A device grouping and pilot set allocation algorithm is proposed. Specifically, devices are divided into multiple groups based on their correlation characteristics. A unique pilot set is assigned to each group. Devices with large channel power overlaps in the angular domain are divided into different groups, since it is difficult to distinguish them if they select the same pilot.
Then the random access protocol in \cite{b4} is utilized. The non-orthogonal pilot hopping pattern over multiple slots is predetermined for devices. The construction of pseudo-random pilot hopping patterns can be modeled as the process that each active device randomly selects a pilot from its pilot set in each slot \cite{b4}.
Since we perform device grouping and pilot set allocation before random access, devices reusing the pilots have less overlapping angle of arrival (AoA) intervals, and impairment caused by pilot interference is reduced. Hence, the proposed scheme shows performance gains over the traditional scheme where devices and pilots are not grouped \cite{b4} in terms of the estimation error and spectral efficiency, and the gains of estimation performance become larger as the channel angular spread (AS) becomes smaller and the signal-to-noise ratio (SNR) becomes higher. Meanwhile, when the channel covariance matrixes of devices reusing the pilots are orthogonal, the estimation error can be minimized.

In this paper, bold lowercase letters and bold uppercase letters denote column vectors and matrices, respectively.
The conjugate, transpose, and conjugate transpose are denoted by $\left(\cdot \right)^{*}$, $\left(\cdot \right)^{T}$, and $\left(\cdot \right)^{H}$, respectively. The Euclidean norm and expectation operators are denoted by $\left\|\cdot \right\|$ and $\mathbb{E}\{\cdot \}$, respectively. Let $z\sim \mathcal{CN}(0,\sigma^{2})$ denote a circularly
symmetric complex Gaussian random variable $z$ with zero mean and variance $\sigma^{2}$.


\section{Channel Model and Channel Estimation}
We consider a massive MIMO system in a single-cell scenario operating in TDD mode, where the BS is equipped with a uniform linear array (ULA) of \emph{M} antennas and serves \emph{K} single-antenna devices. The set of devices within the network is denoted by $\mathcal{K}=\{1,2,\dots,K\}$.

We consider spatially correlated Rayleigh fading channels which are frequency-flat fading on a narrow-band sub-carrier. Let $\textbf{h}_k\in \mathbb{C}^{M\times1}$ $(k\in \mathcal{K})$ denote the UL channel vector between the BS antenna array and device \emph{k}. Let $\theta$ and $\mathcal{A}$ denote the incidence angle and the angular region, respectively. The channel vector $\textbf{h}_k$ can be modeled as \cite{b8,b9}
\begin{equation}
\textbf{h}_k = \int_\mathcal{A}\textbf{v}(\theta)\alpha_k(\theta){\rm d}\theta,
\end{equation}
where $\alpha_k(\theta)$ denotes the channel gain function of device \emph{k}. If BS antennas are spaced with half of wavelength, the steering vector $\textbf{v}(\theta)\!=\![1,e^{-j\pi\sin\theta}\!,\dots,e^{-j\pi(M-1)\sin\theta}]^T$.
Supposing $\textbf{h}_k \!\sim\! \mathcal{CN}(0,\textbf{R}_k)$, the covariance matrix $\textbf{R}_k$ is given by
\begin{equation}
\textbf{R}_k = \int_\mathcal{A} \textbf{v}(\theta)(\textbf{v}(\theta))^Hp_k(\theta){\rm d}\theta,
\end{equation}
where $p(\theta)$ denotes the power azimuth spectrum (PAS), which is assumed to follow the truncated Laplacian distribution in this paper. Let $\varsigma_k$, $\theta_k$, and $\beta_k$ denote the AS, the mean AoA, and the large scale fading coefficient of device \emph{k}, respectively. Then $p_k(\theta)$ equals \cite{b10}
\begin{equation}
p_k(\theta) = \frac{\beta_k {\rm exp}\left({-\sqrt{2}\arrowvert\theta-\theta_k\arrowvert}/{\varsigma_k}\right)}{\sqrt{2}\varsigma_k\left(1-{\rm exp}\left(-\sqrt{2}\pi/\varsigma_k\right)\right)}.
\label{PAS}
\end{equation}

From \cite{b8}, when the number of BS antennas is sufficiently large, the covariance matrix $\textbf{R}_k$ can be approximated by
\begin{equation}
\textbf{R}_k\approx\textbf{F}_M{\rm diag}\{\textbf{r}_k\}\textbf{F}_M^H,
\label{Rk}
\end{equation}
where $\textbf{F}_M$ is a unitary \emph{M}-point Discrete Fourier transform (DFT) matrix. For $i=1,2,\dots,M$, $[\textbf{r}_k]_i$ is given by
\begin{eqnarray}
[\textbf{r}_k]_i = Mp_k\left(\vartheta\left(\frac{i-1}{M}\right)\right)\left[\vartheta\left(\frac{i}{M}\right)-\vartheta\left(\frac{i-1}{M}\right)\right],
\end{eqnarray}
where $\vartheta\left(\frac{m'}{M}\right)\!=\!\arcsin \left(2\frac{m'}{M}-1\right)$ for $m'\!=\!0,1,\dots,M$.
It indicates that when \emph{M} is sufficiently large, channel spatial correlations are related to the channel power distribution in the angular domain. Specifically, eigenvector matrixes of channel covariance matrixes can be approximated by the DFT matrix, and eigenvalues depend on the channel PASs \cite{b8}. Besides, channels are assumed to be wide-sense stationary \cite{b11}, and channel covariance matrixes can be obtained by the BS.

Assuming the pilot is $\tau_p$ symbols long, it is smaller than the channel coherence interval. During the training phase, the set of active devices is denoted by $\mathcal{K}_a$ and its size is assumed to be $K_a$. The pilot of device \emph{k} is denoted by $\pmb{\phi}_{\pi_k}\in \mathbb{C}^{\tau\times1}$, which is the $\pi_k^{\rm th}$ pilot sequence in the pilot set. The transmit power of the pilot signal satisfies $\sigma_{p}^{2}=1$, i.e., $\pmb{\phi}_{\pi_k}^H\pmb{\phi}_{\pi_k}=\tau_p\sigma_{p}^{2}=\tau_p$. Let $\mathcal{C}_{\pi_k}$ denote the set of devices using the same pilot as device \emph{k}.
The received pilot signals at the BS is given by
\begin{equation}
\textbf{Y} = \sum_{l\in \mathcal{K}_a}\textbf{h}_l\pmb{\phi}_{\pi_l}^T+\textbf{N},
\end{equation}
where $\textbf{N}$ is the additive white Gaussian noise (AWGN) whose elements are i.i.d. as $\mathcal{CN}(0,\sigma_{z}^{2})$. The SNR $\rho_p\!=\!\sigma_{p}^{2}/\sigma_{z}^{2}\!=\!1/\sigma_{z}^{2}$.
After decorrelation, the channel observation of device \emph{k} equals
\begin{equation}
\textbf{y}_{\pi_k} = \textbf{Y}\pmb{\phi}_{\pi_k}^{*} = \tau_p\sum_{l\in\mathcal{C}_{\pi_k}}\textbf{h}_l+\textbf{N}\pmb{\phi}_{\pi_k}^*.
\end{equation}

The minimum mean square error (MMSE) estimate of $\textbf{h}_k$ is given by
\begin{equation}
\hat{\textbf{h}}_k = \textbf{R}_k(\sum_{l\in\mathcal{C}_{\pi_k}}\tau_p\textbf{R}_l+\frac{1}{\rho_p}\textbf{I})^{-1}\textbf{y}_{\pi_k}.
\end{equation}

Based on the orthogonality principle of MMSE estimation, the covariance of channel estimation error can be obtained as
\begin{equation}
\textbf{R}_{\tilde{h}_k} = \textbf{R}_k-\tau_p\textbf{R}_k(\sum_{l\in\mathcal{C}_{\pi_k}}\tau_p\textbf{R}_l+\frac{1}{\rho_p}\textbf{I})^{-1}\textbf{R}_k.
\label{MSE}
\end{equation}

\section{DGPSA-based Random Access Protocol} \label{DGPSA-RA}
Since the number of devices within the network is larger than that of orthogonal pilots, it is impossible to allocate a dedicated pilot to each device. Hence, random access becomes a necessary solution.
In this section, a device grouping and pilot set allocation (DGPSA) algorithm is proposed, which is performed before the random access process. The characteristics of correlated channels are fully utilized in the DGPSA-based random access protocol. Finally, the channel estimation error, its theoretical lower bound and the spectral efficiency of the proposed scheme are derived.
\subsection{DGPSA Algorithm}
In this subsection, a DGPSA algorithm is proposed and described in Algorithm \ref{alg:1}. The main idea is that channel covariance matrixes of devices reusing a pilot set should be as orthogonal as possible, i.e., devices with large AoA interval overlaps should be divided into different groups and randomly access to different pilots. Similar ideas were utilized in \cite{b12} to mitigate the inter-cell pilot contamination. However, in this paper Algorithm 1 is dedicated for device grouping and pilot set allocation in multi-user single-cell scenarios.

In Algorithm 1, pilots are equally divided into $Y$ groups as $\mathcal{P}\!=\!\{\mathcal{P}_1,\mathcal{P}_2,\dots,\mathcal{P}_Y\}$. In order to distinguish different devices within a group, the number of pilots in a group should be more than one, which will be illustrated in Section \ref{random p d}.

The process of device grouping consists of two steps.
The first step is to assign devices with similar covariance matrixes to different groups with orthogonal pilot sets. The similarity is measured by the angle between covariance matrixes of different devices. Since these matrixes are Hermitian positive semi-definite, for any two devices \emph{i} and \emph{j} $( i,j\in \mathcal{K}, i\neq j)$, the angle between their covariance matrixes is calculated as
\begin{equation}
\theta(\textbf{R}_i,\textbf{R}_j) = \arccos \frac{{\rm{tr}}\{\textbf{R}_i\textbf{R}_j\}}{\left\|\textbf{R}_i\right\|_{\rm{F}} \left\|\textbf{R}_j\right\|_{\rm{F}} }\in \left[0,\frac{\pi}{2}\right],
\end{equation}
where smaller $\theta$ means weaker orthogonality and stronger similarity.
The second step is to assign each ungrouped device to the group where the channel covariance matrixes of devices are as orthogonal as possible.

The output of Algorithm 1 is the grouping pattern $\mathcal{G}=\{\mathcal{G}_1,\mathcal{G}_2,\dots,\mathcal{G}_Y\}$, which means that devices in group $y$ can randomly access to pilot sequences in the pilot set $\mathcal{P}_y$.

\subsection{Random Access and Channel Estimation}\label{random p d}
Due to random access, collisions will occur in the pilot domain. Hence, it is impossible to distinguish the transmitting devices based on the received pilots in a slot. As a result, random pilot and data access protocol in \cite{b4} is utilized.

\begin{figure}
 \centering
 \includegraphics[width=0.9\linewidth]{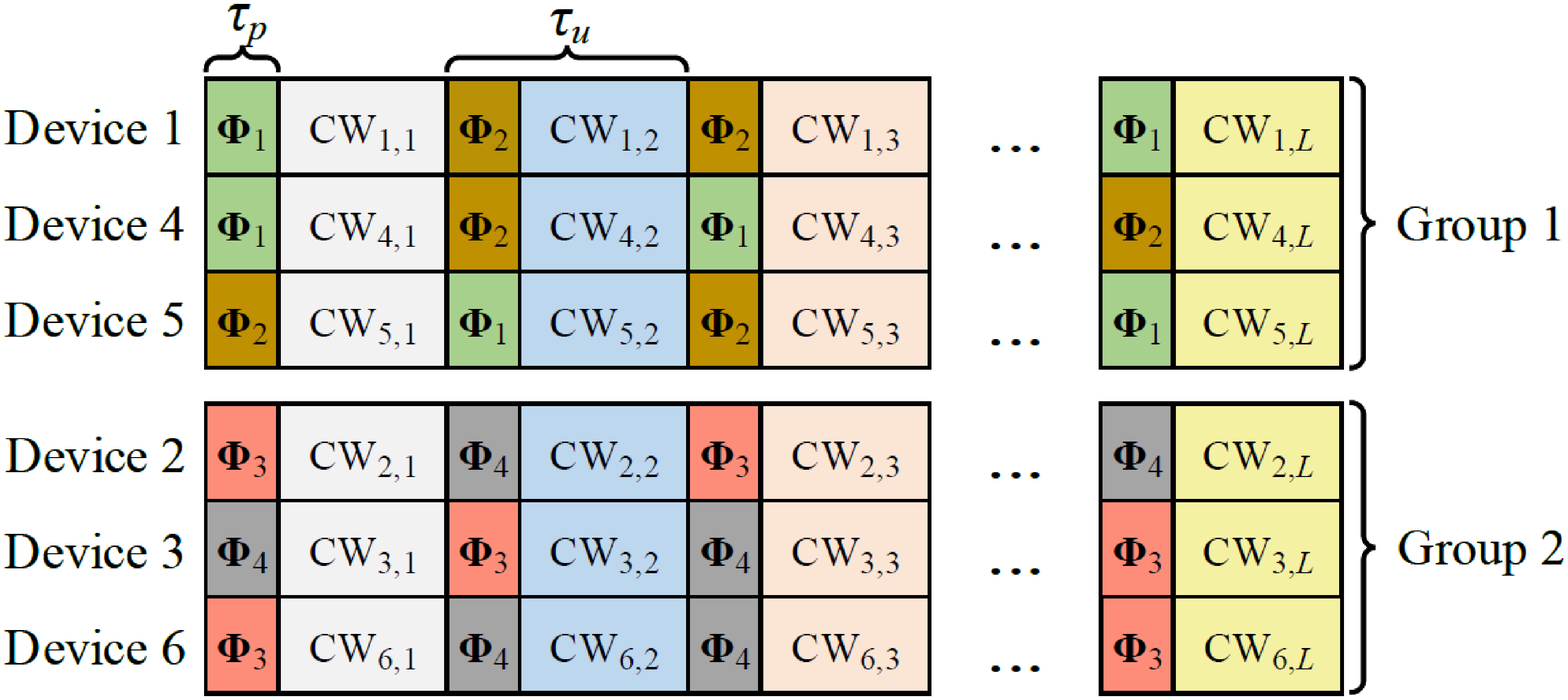}
 \caption{Illustration of the transmission frame. In this example, six devices are divided into two groups. Two orthogonal pilot sets $\{\pmb{\phi}_1,\pmb{\phi}_2\}~{\rm{and}}~\{\pmb{\phi}_3,\pmb{\phi}_4\}$ are assigned to two groups respectively. Each device is allocated with a unique pilot hopping pattern.}
 \label{fig:1}
\end{figure}

Specifically, a UL transmission frame is divided into \emph{L} transmission slots as Fig.~\ref{fig:1}. Each device is associated with a unique and predefined pseudo-random pilot hopping pattern. The pattern of each device consists of pilot sequences in its assigned pilot set. In a transmission slot with $\tau_u$ symbols, active devices select pilot sequences according to their patterns and send part of data codeword afterwards.
At the receiver, the BS runs a correlation decoder across $L$ slots and identifies pilot patterns in order to detect the transmitting devices. Maximum Ratio Combining (MRC) is applied to the codeword and the MRC outputs are combined according to the pilot patterns.

When \emph{L} is large, the transmission of pilots and codewords is affected by asymptotically large number of channel fades and interference events \cite{b4}. Relying on the ergodicity of such a process, the estimation error and spectral efficiency can be characterized as in Section \ref{random p d} and Section \ref{Uplink Sum Rate}.
Since long pilot hopping patterns are used as identifiers, this protocol should be applied to delay-tolerant and low-rate applications.

Assuming the number of devices and pilots in group $y~(1\leq y\leq Y)$ are $U_y$ and $W_y$, respectively, each device can be allocated with a unique non-orthogonal pilot hopping pattern if $(W_y)^L\geq U_y$. Theoretically, when $L$ is sufficiently large, the BS can identify all the devices within the network even if the number of devices is far more than that of pilots.
    \begin{algorithm}[htb]
    \caption{Device Grouping and Pilot Set Allocation (DGPSA) Algorithm}
    \label{alg:1}
    \begin{algorithmic}[1]
    \REQUIRE ~~\\
    The device set $\mathcal{K}=\{1,2,\dots,K\}$;\\
    the channel covariance matrix $\{\textbf{R}_k:k\in\mathcal{K}\}$;\\
    the pilot set $\mathcal{P}=\{\mathcal{P}_1,\mathcal{P}_2,\dots,\mathcal{P}_Y\}$.\\
    \ENSURE ~~\\
    The device grouping pattern $\mathcal{G}=\{\mathcal{G}_1,\mathcal{G}_2,\dots,\mathcal{G}_Y\}$.\\
    \STATE {Initialize the ungrouped device set $\mathcal{K}^{\rm{un}}=\mathcal{K}$, the unused pilot set $\mathcal{P}^{\rm{un}}=\mathcal{P}$}
    \STATE {$\mathcal{G}_1=\{1\},t=2,d_1=1,\mathcal{K}^{\rm{un}}:=\mathcal{K}^{\rm{un}}\backslash\{1\},\mathcal{P}^{\rm{un}}:=\mathcal{P}^{\rm{un}}\backslash\{\mathcal{P}_1\}$}
    \FOR {$t=2$ to $Y$}
    \STATE {Select device $d_t=\mathop {\arg\max}\limits_{i\in \mathcal{K}^{\rm{un}}} \sum_{j\notin\mathcal{K}^{\rm{un}}} \cos\theta(\textbf{R}_i,\textbf{R}_j)$}
    \STATE {Allocate pilot set $\mathcal{P}_t$ to device $d_t$}
    \STATE {Update $\mathcal{K}^{\rm{un}}:=\mathcal{K}^{\rm{un}}\backslash\{d_t\},
    \mathcal{P}^{\rm{un}}:=\mathcal{P}^{\rm{un}}\backslash\{\mathcal{P}_t\},
    \mathcal{G}_t=\{d_t\} $}
    \ENDFOR

    \WHILE {$\mathcal{K}^{\rm{un}}\neq\emptyset$}
    \STATE {Select pilot set $\mathcal{P}_l=\mathop {\arg\min}\limits_{\mathcal{P}_j\in \mathcal{P}} \sum_{s\in\mathcal{K}_{j}} \cos\theta(\textbf{R}_k,\textbf{R}_s)$ for device $k\in\mathcal{K}^{\rm{un}}$}
    \STATE {Allocate pilot set $\mathcal{P}_l$ to device $k$}
    \STATE {Update $\mathcal{G}_{l}:=\mathcal{G}_{l}\cup\{k\},\mathcal{K}^{\rm{un}}:=\mathcal{K}^{\rm{un}}\backslash\{k\}$}
    \ENDWHILE

    \RETURN $\mathcal{G}$
    \end{algorithmic}
    \end{algorithm}

Devices within the network are independently active with the activation probability $p_a$. The sporadic and independent activation of devices and the construction of pseudo-random pilot hopping patterns can be modeled as the process that each active device in each slot randomly selects one of the sequences from its pilot set, and then the probability of having $K_a$ active devices within \emph{K} devices is calculated as \cite{b4}
\begin{equation}
p(K_a|K) = \mathrm{C}_{K}^{K_a}p_a^{K_a}(1-p_a)^{K-K_a}.
\label{K_a}
\end{equation}

Let $\mathcal{U}_{\mathcal{K}_{a}}\!\!\!=\!\!\left\{ \mathcal{U}_{\mathcal{K}_{a}}^1, \mathcal{U}_{\mathcal{K}_{a}}^2,\dots, \mathcal{U}_{\mathcal{K}_{a}}^{N_{\mathcal{K}_{a}}} \!\!\right\}$ denote possible sets of $K_a$ active devices.
Assuming devices in the $l^{\rm {th}}~(1\leq l\leq N_{\mathcal{K}_{a}})$ set are active, its $m^{\rm {th}}~(1\leq m\leq K_a)$ element is denoted by ${\mathcal{K}_{a}^{l,m}}$, and $\mathcal{F}_{\mathcal{K}_{a}^{l,m}}\! = \! \left\{ \mathcal{F}_{{l,m}}^1, \mathcal{F}_{{l,m}}^2,\dots, \mathcal{F}_{{l,m}}^{N_{l,m}} \right\}$ denotes possible collision sets of device ${\mathcal{K}_{a}^{l,m}}$. The number of colliders to device ${\mathcal{K}_{a}^{l,m}}$ is $c$.
Assuming the $n^{\rm {th}}~(1\leq n\leq N_{l,m})$ collision set is considered, the mean square error of channel estimation (MSE-CE) of device ${\mathcal{K}_{a}^{l,m}}$ is given by \cite{b8}
\begin{align}
&~\varepsilon(l,m,n)={\rm{tr}}\left\{\textbf{R}_{\tilde{h}_{\mathcal{K}_a^{l,m},n}}\right\}\notag\\
&=\!{\rm{tr}}\!\left\{\!\textbf{R}_{\mathcal{K}_a^{l\!,m}} \!\!-\!\textbf{R}_{\mathcal{K}_a^{l\!,m}}\!\! \left(\!\textbf{R}_{\mathcal{K}_a^{l\!,m}}\!\!+\!\!\!\!\sum_{\!f\in{\mathcal{F}_{{l\!,m}}^n}}\!\!\!\textbf{R}_{f}\!+\!\frac{1}{\rho_p\tau_p}\textbf{I}\!\right)^{\!\!-1\!}\!\!\!\textbf{R}_{\mathcal{K}_a^{l\!,m}} \!\!\right\}\!\!.
\label{MSE_CE}
\end{align}

Next, we obtain the MSE-CE of a device averaged over all possible sets of $K_a$ active devices, the selection of one device and its colliders. The expected MSE-CE is calculated as
\begin{equation}
\mathbb{E}_{\mathcal{U},\mathcal{K}_a,\mathcal{F}} (\varepsilon)=
\sum\limits_{l=1}^{N_{\mathcal{K}_{a}}}
\sum\limits_{m=1}^{K_{a}}
\sum\limits_{n=1}^{N_{l,m}}
\frac{\varepsilon(l,m,n)} {N_{\mathcal{K}_{a}} K_a N_{l,m}}.
\label{a_MSE_CE}
\end{equation}

Let $U_y$ and $W_y$ denote the number of devices and pilots in group \emph{y}, and $K_{ao}\!=\!K_a\!-1-K\!+U_y$. The probability of having \emph{c} colliders to a given device in group \emph{y} is shown in \eqref{eq_p}.

\newcounter{TempEqCnt} 
\setcounter{TempEqCnt}{\value{equation}} 
\setcounter{equation}{13} 
\begin{figure*}[ht]
\begin{equation}
p(c|K_a) = \left\{ \begin{array}{ll}
\frac{\sum \limits_{j=0}^{K_a-1-c}
\mathrm{C}_{U_y-1}^{c+j}
\mathrm{C}_{c+j}^{c}
{\left(\frac{1}{W_y}\right)}^{c}{\left(1-\frac{1}{W_y}\right)}^{j}
\mathrm{C}_{{K-U_y}}^{K_a-1-c-j}}
{\mathrm{C}_{K-1}^{K_a-1}} & \textrm{if $1\leq K_a\leq U_y,0\leq c\leq K_a-1$}\\
\frac{\sum \limits_{j=0}^{U_y-1-c}
\mathrm{C}_{U_y-1}^{c+j}
\mathrm{C}_{c+j}^{c}
{\left(\frac{1}{W_y}\right)}^{c} {\left(1-\frac{1}{W_y}\right)}^{j}
\mathrm{C}_{{K-U_y}}^{K_a-1-c-j}}
{\mathrm{C}_{K-1}^{K_a-1}} & \textrm{if $U_y<K_a\leq {K-U_y}+1,0\leq c\leq U_y-1$}\\
\frac{\sum \limits_{j=0}^{K-K_a} \mathrm{C}_{U_y-1}^{K_{ao}+j}
\mathrm{C}_{K_{ao}+j}^{c} {\left(\frac{1}{W_y}\right)}^{c} {\left(1-\frac{1}{W_y}\right)}^{K_{ao}+j-c}
\mathrm{C}_{{K-U_y}}^{{K-U_y}-j}}
{\mathrm{C}_{K-1}^{K_a-1}} & \textrm{if ${K-U_y}+1<K_a\leq K,0\leq c\leq K_{ao}$}\\
\frac{\sum \limits_{j=0}^{U_y-1-c}
\mathrm{C}_{U_y-1}^{c+j}
\mathrm{C}_{c+j}^{c}
{\left(\frac{1}{W_y}\right)}^{c} {\left(1-\frac{1}{W_y}\right)}^{j}
\mathrm{C}_{{K-U_y}}^{K_a-1-c-j}}
{\mathrm{C}_{K-1}^{K_a-1}} & \textrm{if ${K-U_y}+1<K_a\leq K,K_{ao}< c\leq U_y-1$}\\
$0$ & \textrm{otherwise.}
\end{array} \right.
\label{eq_p}
\end{equation}
\hrulefill
\end{figure*}
\setcounter{equation}{\value{TempEqCnt}} 

Based on the probability of having $K_a$ active devices in \eqref{K_a} and having $c$ colliders in \eqref{eq_p}, we calculate the expected value of \eqref{a_MSE_CE}, which can be calculated as
\setcounter{equation}{14}
\begin{align}
\overline{\varepsilon} &= p_a \sum\limits_{K_a=1}^{K} p\left(K_a-1|K-1\right)
\sum\limits_{c=0}^{K_a-1} p\left(c|K_a\right)
\mathbb{E}_{\mathcal{U},\mathcal{K}_a,\mathcal{F}} (\varepsilon)\notag\\
&=\sum\limits_{K_a=1}^{K} \frac{K_a}{K} p\left(K_a|K\right)
\sum\limits_{c=0}^{K_a-1} p\left(c|K_a\right)
\mathbb{E}_{\mathcal{U},\mathcal{K}_a,\mathcal{F}} (\varepsilon).
\label{c_a_MSE_CE}
\end{align}

The problem of minimizing $\bar{\varepsilon}$ can be expressed as
\begin{equation}
\begin{split}
&\min~~~\overline{\varepsilon}\\
&\:{\rm{s.t.}} ~~\textbf{R}_k\succeq0,~k\in \mathcal{K}.
\nonumber
\end{split}
\end{equation}

Given the number of devices, $p\left(K_a|K\right)$ is determined for $K_a=1,2,\dots,K$. If $\mathbb{E}_{\mathcal{U},\mathcal{K}_a,\mathcal{F}} (\varepsilon)$ is minimized for all values of $p\left(K_a|K\right)$ and $p\left(c|K_a\right)$, $\overline{\varepsilon}$ can be minimized. Similarly, if $\varepsilon$ is minimized for all kinds of active patterns, all choices of one device and its possible collision events, $\mathbb{E}_{\mathcal{U},\mathcal{K}_a,\mathcal{F}} (\varepsilon)$ can be minimized. Hence, the equivalent problem is given by
\begin{equation}
\begin{split}
&\min~~~\varepsilon(l,m,n)\\
&\:{\rm{s.t.}} ~~\textbf{R}_k\succeq0,~k\in \mathcal{K}~~\:\:~~~ 1\leq l \leq N_{\mathcal{K}_{a}}\\
&~~~~~~~1\leq m \leq K_a~~~~~\:~~~1\leq n \leq N_{l,m}\\
&~~~~~~~1\leq K_a \leq K~~~~~~~~\:0\leq c \leq K_a-1\\
&~~~~~~~l,m,n,K_a,c\in \mathbb{Z}.
\nonumber
\end{split}
\end{equation}

Considering the positive semi-definiteness of the covariance matrix, from \eqref{Rk}, \eqref{MSE} and \eqref{MSE_CE}, we have
\begin{align}
\varepsilon_{\!\rm {min}}(l\!,m\!,n)&\!=\! \sum\limits_{p=1}^M
\left\{r_{\mathcal{K}_a^{l\!,m}}(p)- \frac{\left(r_{\mathcal{K}_a^{l\!,m}}(p)\right)^2}{r_{\mathcal{K}_a^{l\!,m}}(p)+\frac{1}{\rho_p\tau_p}} \right\}\notag\\
&\!=\! {\rm{tr}}\!\left\{\!\textbf{R}_{\mathcal{K}_a^{l\!,m}} \!\!-\!\textbf{R}_{\mathcal{K}_a^{l\!,m}}\!\!\left(\!\textbf{R}_{\mathcal{K}_a^{l\!,m}}\!\!+\!\!\frac{1}{\rho_p\!\tau_p}\textbf{I}\!\right)^{\!\!\!-1\!}\!\!\!\textbf{R}_{\mathcal{K}_a^{l\!,m}}
\!\!\right\}\!\!.
\end{align}
When $\textbf{R}_{\mathcal{K}_a^{l\!,m}} \textbf{R}_{f}\!=\!0$, i.e.,
$\theta\left(\textbf{R}_{\mathcal{K}_a^{l\!,m}}, \textbf{R}_{f}\right)\!=\!\frac{\pi}{2}$ or
$\langle \textbf{r}_{\mathcal{K}_a^{l\!,m}},\textbf{r}_{f} \rangle\!=\!0$ for
$f\!\in\!{\mathcal{F}_{{l\!,m}}^n}$ \cite{b8},
the effect of $\sum_{f\in{\mathcal{F}_{{l\!,m}}^n}}\!\!\textbf{R}_{f}$ can be eliminated and $\varepsilon_{\rm {min}}$ can be obtained. Hence, for any two devices $i$ and $j$ within the same group, if $\textbf{R}_{i} \textbf{R}_{j}\!=\!0$, $\mathbb{E}_{\mathcal{U}\!,\mathcal{K}_{\!a}\!,\mathcal{F}} (\varepsilon)$ can be minimized, which satisfies
\begin{align}
\left[\mathbb{E}_{\mathcal{U},\mathcal{K}_a,\mathcal{F}} (\varepsilon)\right]_{\rm{min}}
&\!\!=\!\mathbb{E}_{\mathcal{U},\mathcal{K}_a,\mathcal{F}} \left(\varepsilon_{\!\rm {min}}\right)\notag\\
&\!\!=\!\frac{1} {K}\! \sum\limits_{i\in \mathcal{K}}\!{\rm{tr}}\!\left\{\!\textbf{R}_{i} \!-\!\textbf{R}_{i}\!\left(\!\textbf{R}_{i}\!+\!\frac{1}{\rho_p\!\tau_p}\textbf{I}\!\right)^{\!\!\!-1}\!\!\!\textbf{R}_{i}
\!\right\}\!.
\end{align}
$\left[\mathbb{E}_{\mathcal{U},\mathcal{K}_a,\mathcal{F}} (\varepsilon)\right]_{\rm{min}}$ does not vary with the fluctuation of $p\left(c|K_a\right)$. Considering $\sum_{c=0}^{K_a-1} \!p\left(c|K_a\right)\!=\!1$, the theoretical minimum value of $\overline{\varepsilon}$ is given by
\begin{align}
\overline{\varepsilon}_{\rm{min}}
&\!=\!\!\sum\limits_{K_a=1}^{K} \frac{K_a}{K} p\left(K_a|K\right) \sum\limits_{c=0}^{K_a-1} p\left(c|K_a\right) \left[\mathbb{E}_{\mathcal{U},\mathcal{K}_a,\mathcal{F}} (\varepsilon)\right]_{ {\rm{min}}}\notag\\
&\!=\!\!\sum\limits_{K_a=1}^{K} \!\frac{K_a}{K} p\left(K_a|K\right) \left[\mathbb{E}_{\mathcal{U},\mathcal{K}_a,\mathcal{F}} (\varepsilon)\right]_{\rm{min}}  \notag\\
&\!=  p_a \left[\mathbb{E}_{\mathcal{U},\mathcal{K}_a,\mathcal{F}} (\varepsilon)\right]_{\rm{min}} .
\end{align}

Hence, for each device, if its channel covariance matrix is orthogonal to that of its colliders, the estimation error can be minimized. This can be realized when devices sharing a pilot set have orthogonal covariance matrixes, i.e., they have non-overlapping AoA intervals. The purpose of Algorithm 1 is to make devices with approximately orthogonal matrixes within a group and make the interference $\sum_{\!f\in{\mathcal{F}_{{l\!,m}}^n}}\!\!\!\textbf{R}_{f}$ as limited as possible.
Besides, given the number of pilots in a group, when the pilot length decreases, the number of groups decreases, i.e., the number of possible colliders to a device increases and thus $\sum_{\!f\in{\mathcal{F}_{{l\!,m}}^n}}\!\!\!\textbf{R}_{f}$ increases. Hence, the MSE-CE gap between the DGPSA-based random access scheme and the theoretical lower bound decreases as the pilot length increases.

\subsection{Uplink Sum Rate}\label{Uplink Sum Rate}
Assuming $K_a$ devices in set $\mathcal{U}_{\mathcal{K}_{a}}^l~(1\leq l\leq N_{\mathcal{K}_{a}})$ are active, during the data transmission phase, the signal $s_{\mathcal{K}_{a}^{l,i}}$ is transmitted from device $\mathcal{K}_{a}^{l,i}\in \mathcal{U}_{\mathcal{K}_{a}}^l$ to the BS antenna array, where $s_{\mathcal{K}_{a}^{l,j}} \!\sim \!\mathcal{CN}\left(0,1\right)$. The received signal is given by
\begin{equation}
\textbf{Y}^u=\sum_{s_{\mathcal{K}_{a}^{l,i}}\in \mathcal{U}_{\mathcal{K}_{a}}^l} \textbf{h}_{\mathcal{K}_{a}^{l,i}} s_{\mathcal{K}_{a}^{l,i}}+\frac{1}{\sqrt{\rho^u}}\textbf{n}^u,
\end{equation}
where $\textbf{h}_{\mathcal{K}_{a}^{l,i}} \in \mathbb{C}^{M\times 1}$ is the channel vector of active device ${\mathcal{K}_{a}^{l,i}}$, $\rho^u$ is the data transmission SNR, and $\textbf{n}^u \sim \mathcal{CN}\left(\textbf{0},\textbf{I}_M \right)$ is the independent additive noise.

The possible collision events of active devices in set $\mathcal{U}_{\mathcal{K}_{a}}^l$ are denoted by $\mathcal{Q}_{{l}}\!=\!\left\{ \mathcal{Q}_{{l}}^1, \mathcal{Q}_{{l}}^2, \dots, \mathcal{Q}_{{l}}^{N_{\mathcal{Q}}}\right\}$. The $q^{\rm{th}}~(1\leq q\leq {N_{\mathcal{Q}}})$ possible collision event is considered. The MRC is utilized at the BS, i.e., $\textbf{v}_{\mathcal{K}_{a}^{l,m}\!,q} \!= \hat{\textbf{h}}_{\mathcal{K}_{a}^{l,m}\!,q}$ \cite{b13}.
We rewrite `` $\mathcal{K}_{a}^{l,m},q$'' to ``$lmq$'' for short, and obtain
\begin{align}
\textbf{v}_{lmq}^{\rm{H}}\textbf{Y}^u=&~\textbf{v}_{lmq}^{\rm{H}} \hat{\textbf{h}}_{lmq} s_{\mathcal{K}_{a}^{l,m}} + \textbf{v}_{lmq}^{\rm{H}} \tilde{\textbf{h}}_{lmq} s_{\mathcal{K}_{a}^{l,m}}\notag\\
&+\!\! \!\!\sum\limits_{j=1,j\neq m}^{K_a} \!\!\!\! \textbf{v}_{lmq}^{\rm{H}} \textbf{h}_{ljq} s_{\mathcal{K}_{a}^{l,j}} + \frac{1}{\sqrt{\rho^u}} \textbf{v}_{lmq}^{\rm{H}} \textbf{n}^u.
\end{align}

The spectral efficiency of device ${\mathcal{K}_{a}^{l,m}}$ is given by \cite{b13}
\begin{align}
&{\rm {SE}}(l,m,q)\!=\! \frac{\tau_u\!-\!\tau_p}{\tau_u}\notag\\
&\mathbb{E}\! \left\{\! {\rm{log}}_2\!\! \left(\! \! 1\!+\! \frac{\arrowvert \textbf{v}_{lmq}^{\rm{H}} \hat{\textbf{h}}_{lmq} \arrowvert^2} {\!\!\textbf{v}_{lmq}^{\rm{H}} \! \left( \! \sum\limits_{j\neq m}\! \hat{\textbf{h}}_{ljq} \hat{\textbf{h}}_{ljq}^{\!{\rm{H}}}\!\! +\!\! \sum\limits_{j=1}^{K_a} \textbf{R}_{\tilde{\textbf{h}}_{ljq}} \! \!\!+ \! \!\frac{1}{\rho^u}\textbf{I} \! \right) \! \textbf{v}_{lmq}} \!\right)\!\! \right\}\!\!.
\label{SE}
\end{align}
Similar to the process in Section \ref{random p d}, we calculate the expected value of \eqref{SE} with respect to the number of active devices, active patterns, the number of collision devices, and collision events of active devices. Finally, the expected spectral efficiency of devices within the network is given by
\begin{equation}
\begin{split}
\overline{\rm {SE}}\!=\!\!\!\sum\limits_{K_{\!a}\!=1}^{K} \!\! K_a p\!\left(\!K_a|K\!\right) \!\!
\sum\limits_{c=0}^{K_{\!a}\!-1}\!\! p\!\left(c|K_a\!\right) \!
\sum\limits_{l=1}^{N_{\!\mathcal{K}_{\!a}}}\! \sum\limits_{m=1}^{K_{\!a}} \! \sum\limits_{q=1}^{N_{\!\mathcal{Q}}} \!\frac{{\rm {SE}}(l\!,m\!,q)} {N_{\mathcal{K}_{\!a}} \!K_a N_{\!\mathcal{Q}}}\!.\!
\end{split}
\label{P_SE}
\end{equation}

\section{Simulation Results}
In this section, we present the simulation and analysis results to evaluate the performance of the proposed scheme.

We consider the UL system with $120$ devices, where the BS is equipped with the 128-antenna ULA spaced with a half wavelength. We use the truncated Laplacian distribution in \eqref{PAS} to generate the channel PAS, and the channel power is normalized \cite{b8}. For devices within the network, we assume their angular spread degrees (ASDs) are equal, i.e., $\varsigma\!=\!\varsigma_k$ for $k\!\in\! \mathcal{K}$, and their large scale fading coefficients are assumed to be 1. We assume their mean channel AoAs are uniformly distributed within the interval $\left[-\frac{\pi}{3},\frac{\pi}{3} \right]$. The channel SNR in training phase and data transmission phase are equal.

\begin{figure}
  \centering
  \includegraphics[width=0.95\linewidth]{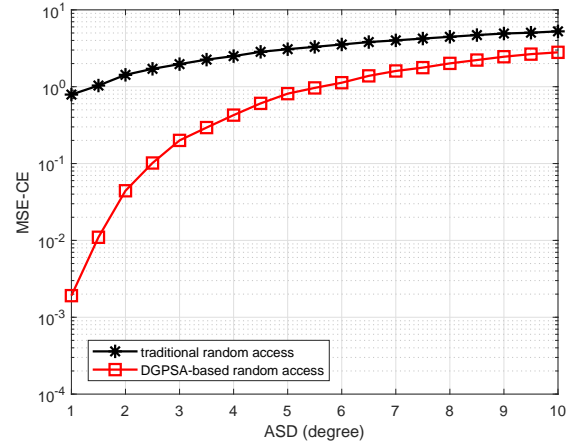}\\
  \caption{Comparison of the MSE-CE performances between the DGPSA-based random access scheme and traditional random access scheme where devices and pilot sets are not grouped. Results are shown versus ASD with $K=120$, $p_a=1/3$, and $\tau_p=40$.}
  \label{fig:2}
\end{figure}
We employ the MSE-CE metric to evaluate performances of the DGPSA-based random pilot and data access scheme developed in Section \ref{DGPSA-RA}. In Fig.~\ref{fig:2}, $40$ pilots are equally divided into $20$ groups and $120$ devices are divided based on Algorithm 1. Since the number of devices in different groups is approximately equal, the assignment of pilots is reasonable.
Assuming $p_a\!=\!\frac{1}{3}$ and $\rho_p\!=\!\rho_u\!\!=\!20$ dB, Fig.~\ref{fig:2} shows the MSE-CE of the DGPSA-based random access scheme and the traditional random access scheme \cite{b4}. In \cite{b4}, devices and pilots are not grouped.
It can be observed that the proposed scheme outperforms the traditional scheme in terms of the MSE-CE in all ASD regime. This is because the overlapping AoA intervals of possible colliders in this paper is less than that in \cite{b4}. Especially, if the ASD is small, i.e., channels are strongly correlated, the improvement is significant.

\begin{figure}
  \centering
  \includegraphics[width=0.95\linewidth]{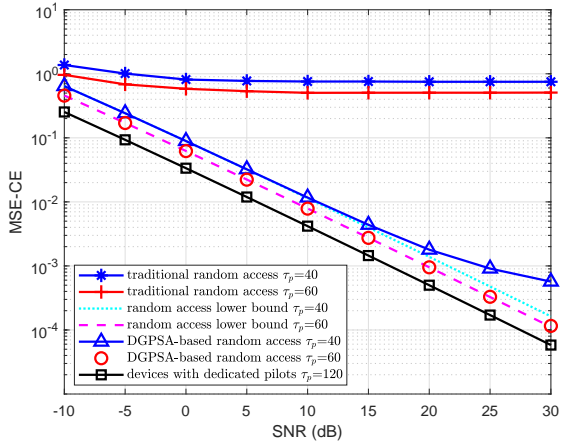}\\
  \caption{Comparison of the MSE-CE performances between DGPSA-based random access scheme, its theoretical lower bound, traditional random access scheme, and the ideal case where devices have dedicated pilots. Results are shown versus SNR with $K=120$, $p_a=1/3$, and $\varsigma=1^{\circ}$.}
  \label{fig:3}
\end{figure}
\begin{figure}
  \centering
  \includegraphics[width=0.95\linewidth]{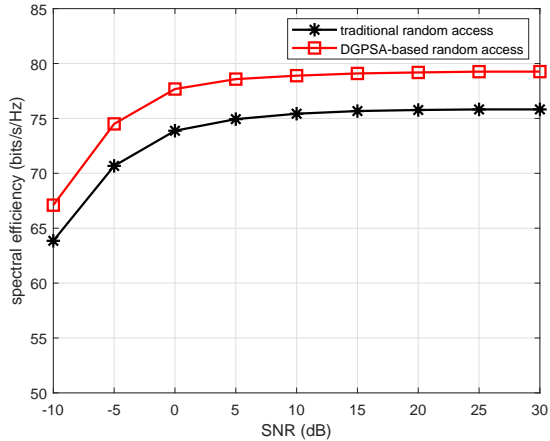}\\
  \caption{Comparison of the spectral efficiency between the DGPSA-based random access scheme and traditional scheme. Results are shown versus SNR with $K=120$, $p_a=1/2$, $\tau_u = 128$, $\tau_p = 30$, and $\varsigma=2^{\circ}$.}
  \label{fig:4}
\end{figure}
In Fig.~\ref{fig:3}, for the DGPSA-based random access scheme, pilots and devices are divided into $\frac{\tau_p}{2}$ groups based on Algorithm 1. When $\tau_p\!=\!120$, each device is pre-allocated with a dedicated pilot. Assuming $p_a\! =\!\frac{1}{3}$ and $\varsigma \!=\! 1^\circ$, Fig.~\ref{fig:3} shows the MSE-CE of the proposed scheme, its theoretical lower bound, traditional random access scheme \cite{b4}, and the ideal case where devices have dedicated pilot sequences.
We observe from Fig.~\ref{fig:3} that the MSE-CE improves as the pilot length increases. Besides, the performance of the proposed scheme is close to the theoretical lower bound regardless of the pilot length in low SNR regime where noise influence dominates. When the SNR is high and pilot interference dominates, the MSE-CE performance gap between the proposed scheme and its theoretical lower bound increases as the pilot length decreases. This is because the proposed algorithm makes the possible collision devices have approximately orthogonal channel covariance matrixes but not strictly orthogonal matrixes, and shorter pilot will increase the number of colliders and the residual interference. However, the proposed scheme shows obvious performance gains over the traditional scheme especially in high SNR regime.
Furthermore, the performance gap between the proposed scheme with $\tau_p\!=\!60$ and the ideal case with $\tau_p\!=\!120$ is not large.

In Fig.~\ref{fig:4}, when $K\!=\!120$, $p_a\!=\!1/2$, $\tau_p\!=\!30$, $\tau_u\!=\!128$, and $\varsigma\!=\!2^{\circ}$, the spectral efficiencies of the proposed scheme and traditional random access scheme \cite{b4} are presented. It is shown that the spectral efficiency of the proposed scheme is higher than that of the traditional scheme in all SNR regime due to the reduce of MSE-CE.

\section{Conclusion}
Random access has been an important topic because the number of pilot sequences is limited and the activities of devices are sporadic. However, most of related works concentrate on i.i.d. channels with fewer focus on realistic correlated outdoor wireless propagation environments.
In this work, a DGPSA-based random pilot and data access protocol is proposed for massive MIMO systems with spatially correlated Rayleigh fading channels. Specifically, devices and pilot sets are divided into different groups according to the DGPSA algorithm, and devices within the same group have less overlapping channel AoA intervals. Then active devices perform random pilot and data access. The theoretical MSE-CE lower bound is derived.
The simulation results show that the proposed scheme outperforms the traditional scheme in terms of the MSE-CE and spectral efficiency. Furthermore, the MSE-CE performance gains are more significant in smaller ASD and higher SNR regime. Besides, the MSE-CE of the proposed scheme is close to its theoretical lower bound over a wide SNR region especially for long pilot sequence.
Hence, the DGPSA-based random pilot and data access protocol is crucial and it is suitable to multiple low-power and intermittently active devices in massive MIMO systems with spatially correlated Rayleigh fading channels.




\begin{thebibliography}{00}
\bibitem{b1} N. G. M. N. Alliance, ``5G white paper,'' Next Generation Mobile Networks, White Paper, pp. 1--125, 2015.
\bibitem{b2} C. Bockelmann, N. Pratas, H. Nikopour, K. Au, T. Svensson, C. Stefanovic, P. Popovski, and A. Dekorsy, ``Massive machine-type communications in 5G: Physical and MAC-layer solutions,'' \emph{IEEE Commun. Mag.}, vol. 54, pp. 59--65, Sep. 2016.
\bibitem{b3} T. L. Marzetta, ``Noncooperative cellular wireless with unlimited numbers of base station antennas,'' \emph{IEEE Trans.  Wireless Commun.}, vol. 9, pp. 3590--3600, Nov. 2010.
\bibitem{b4} E. de Carvalho, E. Bj\"ornson, J. H. S\o rensen, E. G. Larsson, and P. Popovski, ``Random pilot and data access in massive MIMO for machine-type communications,''\emph{ IEEE Trans. Wireless Commun.}, vol. 16, pp. 7703--7717, Dec. 2017.
\bibitem{b5} E. de Carvalho, E. Bj\"ornson, J. H. S\o rensen, P. Popovski, and E. G. Larsson, ``Random access protocols for massive MIMO'', \emph{IEEE Commun. Mag.}, vol. 55, pp. 216--222, May 2017.
\bibitem{b6} J. H. S\o rensen, E. de Carvalho, and P. Popovski, ``Massive MIMO for crowd scenarios: A solution based on random access,'' in \emph{IEEE Globecom Workshops} (\emph{GC Wkshps}), Austin: Academic, Dec. 2014, pp. 352--357.
\bibitem{b7} E. Bj\"ornson, E. de Carvalho, J. H. S\o rensen, E. G. Larsson, and P. Popovski, ``A random access protocol for pilot allocation in crowded massive MIMO systems,'' \emph{IEEE Trans. Wireless Commun.}, vol. 16, pp. 2220--2234, Apr. 2016.
\bibitem{b8} L. You, X. Q. Gao, X. G. Xia, N. Ma, and Y. Peng, ``Pilot reuse for massive MIMO transmission over spatially correlated Rayleigh fading channels,'' \emph{IEEE Trans. Wireless Commun.}, vol. 14, pp. 3352--3366, Feb. 2015.
\bibitem{b9} X. Meng, X. Q. Gao, and X. G. Xia, ``Omnidirectional precoding based transmission in massive MIMO systems,'' \emph{IEEE Trans. Commun.}, vol. 64, pp.174--186, Nov. 2016.
\bibitem{b10} K. I. Pedersen, P. E. Mogensen, and B. H. Fleury, ``A stochastic model of the temporal and azimuthal dispersion seen at the base station in outdoor propagation environments,'' \emph{IEEE Trans. Veh. Technol.}, vol. 49, pp. 437--447, Mar. 2000.
\bibitem{b11} B. Clerckx, and C. Oestges, \emph{MIMO Wireless Networks: Channels, Techniques and Standards for Multi-Antenna, Multi-User and Multi-Cell
Systems}, 2nd ed. Oxford, UK: Academic Press, 2013.
\bibitem{b12} H. F. Yin, D. Gesbert, M. Filippou, and Y. Z. Liu, ``A coordinated approach to channel estimation in large-scale multiple-antenna systems,''\emph{ IEEE J. Sel. Area Commun.}, vol. 31, pp. 264--273, Feb. 2013.
\bibitem{b13} E. Bj\"ornson, J. Hoydis, and L. Sanguinetti, \emph{Massive MIMO Networks: Spectral, Energy, and Hardware Efficiency}, Foundations and Trends$^{\textcircled{\scriptsize{R}}}$ in Signal Processing: vol. 11, pp. 154--655, Nov. 2017.

\end{thebibliography}
\end{document}